# L'éthique des systèmes d'information autonomes vers une pensée artificielle

Joël Colloc

Laboratoire IDEES Le Havre UMR CNRS 6266, Université du Havre,
Normandie Université
joel.colloc@univ-lehavre.fr,
WWW home page: **http://www.cirtai.org/spip.php7rubrique389**

**Résumé :** Des projets, situés au carrefour des sciences cognitives, des neurosciences, de l'informatique et de la robotique concernent actuellement la création d'êtres artificiels autonomes capables de penser. Cet article présente un modèle de comparaison de la pensée humaine avec une hypothétique pensée numérique fondée sur quatre hiérarchies : la classification des systèmes d'information, la pyramide cognitive, la pyramide langagière et la hiérarchie informationnelle numérique. Après un état de l'art sur la nature de la pensée humaine, la faisabilité de systèmes multi-agents autonomes dotés d'une conscience artificielle et capable de penser est discutée. Les aspects éthiques et les conséquences pour l'humanité de tels systèmes sont évalués. Ils conduisent actuellement à une réaction de la communauté scientifique.

**Mots-clés :** éthique, pensée artificielle, holisme, systèmes multi-agents autonomes, conscience artificielle

**Title :** The ethics of autonomous information systems towards an artificial thinking

**Abstract :** Projects, situated in the crossroads of the cognitive sciences, the neurosciences, the computing and robotics currently concern the creation of autonomous artificial beings capable of thinking. This article presents a model of comparison of the human thinking with a numerical thinking that relies on four hierarchies: the classification of information systems, the cognitive pyramid, the linguistic pyramid and the numerical information hierarchy. After a state of the art on the nature of the human thinking, the feasibility of autonomous multi-agents systems endowed with an artificial consciousness and capable of thinking is discussed. The ethical aspects and the consequences for the humanity of such systems are evaluated. This subject causes at present the reaction of the scientific community.

## 1 Introduction

Les progrès récents des sciences cognitives, des neurosciences, de l'informatique et de la robotique relancent le projet de la création d'un être artificiel autonome capable de penser. Cet article étudie la faisabilité d'un tel projet en comparant la manière dont les humains acquièrent, exploitent leurs informations et élaborent la pensée avec les capacités actuelles des systèmes d'information autonomes. Notre comparaison est fondée sur quatre hiérarchies : la hiérarchie des systèmes d'information, issue de la systémique, fournit un indicateur de complexité et d'autonomie; la hiérarchie cognitive décrit l'acquisition sub-symbolique et l'émergence de notre expérience personnelle, notre savoir, tandis que la hiérarchie langagière construit le discours décrivant les connaissances acquises en termes d'objets concrets et abstraits sur l'environnement et sur soi-même. La hiérarchie informationnelle numérique s'inscrit dans l'histoire de l'informatique avec l'augmentation considérable de la puissance des ordinateurs et de l'intelligence artificielle qui a produit des concepts, des modèles, des méthodes et des outils nécessaires à la réalisation de systèmes d'information autonomes. Deux questions essentielles se posent à laquelle nous tentons de répondre la première : un système autonome est-il capable de générer une forme de pensée artificielle ? La seconde quelles sont les retombées éthiques et épistémologiques de la réalisation de systèmes dotés d'une pensée artificielle numérique ? Cette réflexion intervient dans le contexte où des architectures de tels systèmes existent déjà et des recherches sur le "transhumanisme" sont actuellement effectuées.

## 2  La hiérarchie des systèmes d'information

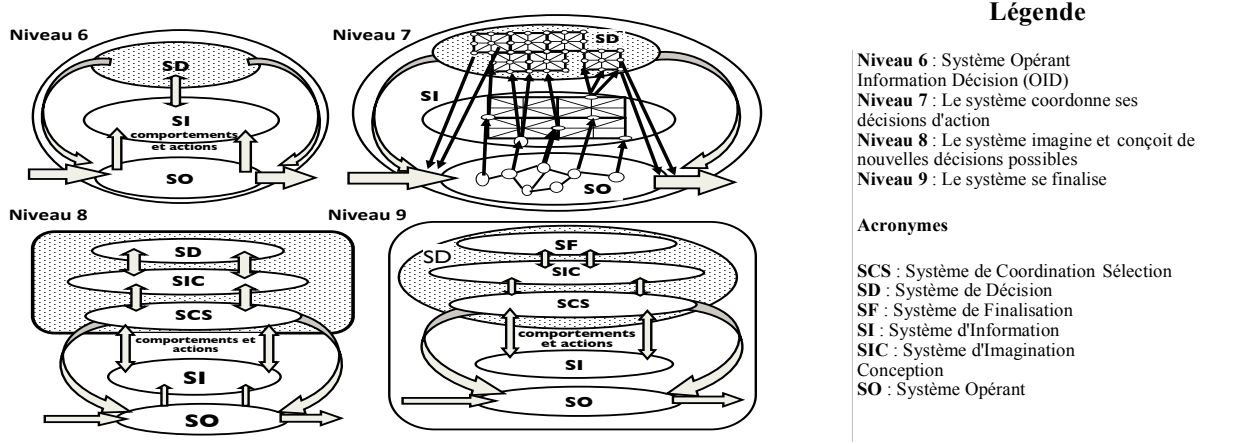

Fig. 1: Les niveaux 6 à 9 du modèle de système complexe de J-L. Le Moigne© [45]

La théorie générale des systèmes est très utilisée en sciences humaines, notamment en sociologie et en gestion pour modéliser les organisations, qui peuvent être considérées comme des êtres vivants sociaux, et les systèmes d'information dont les flux d'information constituent la véritable circulation sanguine de ces organismes. La classification des systèmes proposée par Jean-Louis Le Moigne inspirée de K. Boulding décrit l'évolution des systèmes d'information. Il fonctionne comme une épigenèse : le niveau suivant ajoutant des structures et des fonctionnalités au niveau précédent. Par concision, nous ne rappelons que les niveaux 6 à 9 utilisés dans cet article et représentés sur la figure 1. Les précédents 1 à 5 peuvent être consultés dans [45]. Sur la figure 1 niveau 6 : Le système est doté d'un sous-système de mémorisation pour la décision. Ce modèle O.I.D (Système Opérant (**SO**), Système d'Information (**SI**), Système de Décision (**SD**)) est considéré comme canonique car fondateur d'un grand nombre de théories des systèmes d'information. Les trois sous-systèmes peuvent être décomposés par récursivité en sous-systèmes spécialisés stables du système de décision. Sur la figure 1 niveau 7 : Les opérations du système de décision consistent à coordonner de nombreuses décisions d'actions que le système doit prendre à chaque instant t, en tenant compte des informations endogènes (sa propre activité et régulation) et exogènes : les informations provenant de l'environnement (entrées) et ses actions décidées pour tenter d'agir sur l'environnement (sorties). Ils correspondent aux systèmes multi-agents distribués capables de coordonner leurs décisions dans un réseau complexe d'agents partenaires en vue d'établir une intelligence artificielle distribuée et collective. Sur la figure 1 niveau 8: Le système devient capable d'élaborer de nouvelles formes d'actions à l'aide de son imagination. Pour rendre compte de cet aspect, le modélisateur doit pouvoir faire émerger un sous-système d'imagination et de conception (**SIC**) de nouvelles formes ou aptitudes de décision. Sur la figure 1 niveau 9 : Le système est parfois capable de décider sur sa décision et de déterminer les aspects positifs et négatifs de ses actions passées sur l'environnement. Certains systèmes naturels ou artificiels peuvent ainsi se finaliser. L'hypothèse de finalisation d'un système complexe est ainsi posée par Jean-Louis Le Moigne. Il s'agit d'une forme de méta-connaissance où le système s'interroge sur le bien fondé de ses propres décisions, actions et leurs résultats. Cette activité est proche de la pensée humaine (SF) qui lui confère une autonomie de décision avec lequel il peut ainsi prendre en charge son destin et fixer ses propres objectifs. Une approche analytique de cette hypothèse, conforme à l'empirisme logique, bien que recevable ne trouve pas de réponse à l'étude de tels systèmes. L'approche systémique est la seule voie praticable pour appréhender la complexité des systèmes vivants. Ils correspondent aux systèmes multi-agents autonomes.

## 3  Une double hiérarchie de l'information humaine

### 3.1  Les voies sensorielles et l'émergence des objets

Les cinq sens les plus connus sont la vision, l'ouie, l'odorat, le goût et le toucher mais ils sont en fait beaucoup plus nombreux. Le toucher comprend en fait un grand nombre de capteurs cutanés sensibles à la pression (bari-récepteurs), à la température, à la douleur, aux vibrations. La sensibilité proprioceptive nous renseigne sur la position, l'orientation de nos membres, de notre corps dans l'espace. Le système vestibulaire de l'oreille interne régule notre équilibre... A côté des voies sensorielles comme la vision, certaines aires associatives sont spé-cialisées dans l'analyse des visages situées dans le lobe temporal [58] ou les mouvements dans l'aire temporale moyenne (MT) [43]. La neurophysiologie des voies sensorielles est décrite avec détails dans [2,36,58,59]. Pour toutes les voies sensorielles on retrouve globalement une organisation similaire ascendante : des capteurs périphériques, une transduction en signal électrique, la transmission (+/- longue) par les nerfs dans le tronc cérébral ou par des paires de nerfs crâniens (V,VII, IX, X), le relai dans une aire thalamique spécifique, une projection dans une zone corticale spécialisée et des aires associatives communes à plusieurs modalités sensorielles. A chaque étage, il existe de nombreux neurones récurrents formant une "rétropropagation" qui assurent les boucles de rétroaction essentielles au fonctionnement du cortex cérébral fondé sur la prédiction des futures perceptions attendues comme cohérentes avec celles qui viennent d'être mémorisées et ce à chaque niveau de détails : des éléments de perception éphémères, que nous appelons *"percepts"* jusqu'aux objets mémorisés durablement figure 3. La description de ces connexions récurrentes n'a été que très tardive et leur rôle a été longtemps mal compris. Nous disposons d'un système sensoriel global : la vue, l'ouïe, le toucher et le goût et l'odorat sont combinés, les flux montent et descendent le long d'une hiérarchie à branches multiples et les prédictions sensorielles potentielles ne sont pas quelconques mais reposent sur celles qui ont déjà été mémorisées dans les différentes couches du cortex [37] figure 2.

### 3.2 L'intégration des voies sensorielles et de la motricité

La motricité est représentée sur un modèle similaire par des voies descendantes et sont fortement intriquées avec les voies sensorielles et disposent également de boucles de rétro-action. Vernon Benjamin Mountcastle a formulé l'hypothèse que la structure et le fonctionnement du cortex cérébral était finalement assez uniforme sur l'ensemble de sa surface et que le cortex moteur ressemble au cortex sensoriel. Le cortex traite les prédictions sensorielles ascendantes de manière analogue aux commandes motrices descendantes [37,50].

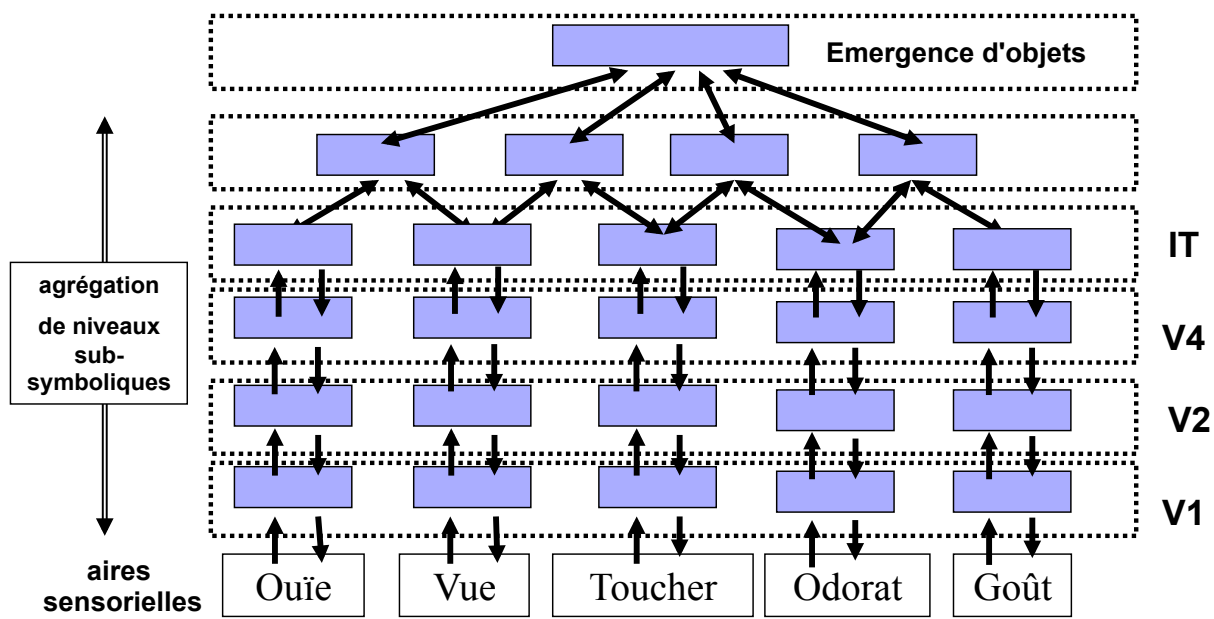

Fig. 2: Emergence sensorielle d'objets connus [37]

De plus, l'intégration des sens a lieu parce qu'il existe de nombreuses aires associatives vers laquelle convergent les différentes voies sensorielles ascendantes et des boucles de rétroactions qui produisent des prédictions non seulement pour le sens considéré mais aussi pour tous les autres. Par exemple le fait de voir une pomme active l'odeur du fruit, sa forme, sa consistance si je dois la toucher (en relation avec la motricité) et le mot pomme si je dois en parler. Le fait de sentir l'odeur de pomme (les yeux fermés) va activer une image stéréotypée de pomme que l'on s'attendrait à voir. L'ensemble de ces prédictions est possible car nous mémorisons une perception globale unifiée de l'objet pomme reliée aux différentes autres modalités sensorielles, motrices, langagières qui lui correspondent [37]. Evoquer son nom permet d'imaginer, sa forme son odeur, sa consistance [49]. La figure 2 montre les différentes couches corticales V1,V2,V4 qui intègrent les différentes modalités sensorielles et les zones associatives du cortex inféro-temporal (IT) qui s'activent quand l'image d'une pomme est reconnue ou lorsque l'odeur d'une pomme est perçue ou encore si le mot pomme est prononcé. Cette activation est alors transmise de manière descendante aux autres aires sensorielles ou langagières qui vont prédire les patterns correspondants à des niveaux de détails de plus en plus fin pour confirmer ou infirmer qu'ils sont concordants dans chacun des autres sens. Nous pouvons ainsi savoir si nous sommes en présence d'une vraie pomme ou seulement d'une représentation partielle : comme dans le fameux tableau de Magritte qui porte l'inscription "ceci n'est pas une pomme".

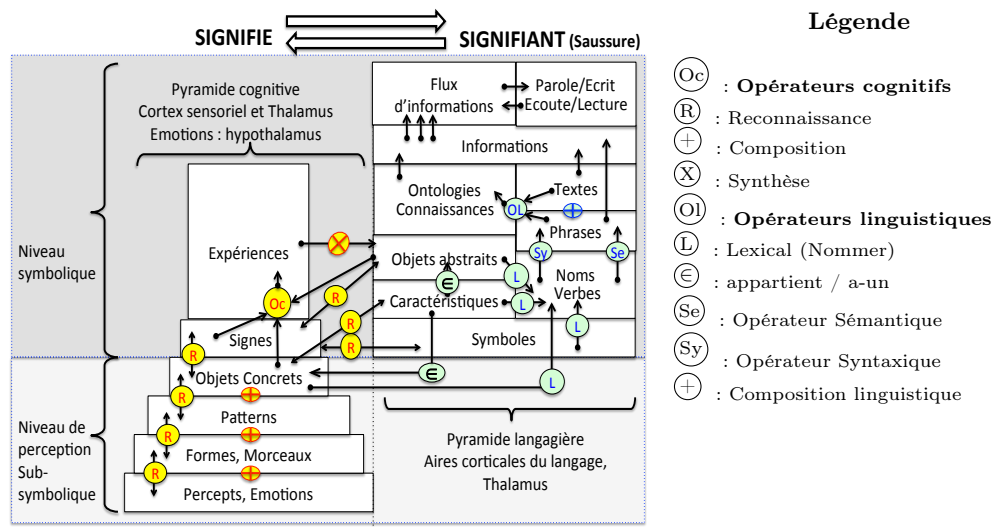

Fig. 3: Double Hiérarchie cognitive / langagière

A cette organisation anatomo-physiologique correspond une hiérarchie des informations sensorielles transmises et intégrées au niveau cortical décrite sur la figure 3. L'hippocampe est la zone située au sommet de la pyramide mnésique. Cette structure cérébrale est à l'origine de l'appariement et l'interconnexion des différentes modalités sensorielles et la mémorisation des caractéristiques des nouveaux objets inconnus [37].

## 4 Une double hiérarchie informationnelle chez l'homme

### 4.1 Un modèle unificateur du système d'information humain

La figure 3 propose un modèle conceptuel du système d'information humain. Dans sa partie gauche on trouve la pyramide sensorielle et cognitive conduisant à l'émergence d'objets concrets et à la mémorisation de connaissances et d'expériences sur les objets du monde conformément à la section précédente. Les opérateurs cognitifs sont : 0 qui représente la composition de percepts en formes (ou morceaux) en patterns puis en objets concrets. A chaque niveau des opérateurs de reconnaissance (R) confirment ou infirment la cohérence dans le flux des perceptions. Les percepts sont des éléments primaires perçus par les organes des sens. Pour la vision, il s'agit de points, de segments orientés (verticaux, obliques, horizontaux...) qui constituent dans le champ visuel des formes élémentaires comme celles décrites dans la théorie des *"géons"* proposée par Hummel et Biederman en 1987 [4,5,36]. La combinaison de ces formes élémentaires *"géons"* compose des patterns visuels caractéristiques d'objets concrets mémorisés dans le cortex visuel et associés dans les aires d'association avec les autres modalités sensorielles et les aspects émotionnels (hypothalamiques) les concernant [22,23].

### 4.2 Les émotions liées à la perception

Le cerveau est aussi une glande dont les neurohormones, sécrétées principalement par l'hypothalamus et l'amygdale, influencent l'interprétation des informations sensorielles et interagissent avec les autres parties du système nerveux. L'hypothlamus régule l'ensemble des équilibres du corps (faim & soif/satiété, plaisir/douleur, ...) [64,65]. L'amygdale intervient dans les circuits de la récompense et de la motivation. L'amygdale déclenche des réponses au stress à des stimuli menaçants en stimulant l'hypothalamus qui libère une hormone CRH (corticotropin releasing Hormon) stimulant la sécrétion par l'hypophyse d'ACTH (Adenocorticortrophine hormon) qui va stimuler la sécrétion de cortisol par les glandes surrénales situées sur les reins [46]. Ils forment un système indissociable avec le système nerveux périphérique, le système sympathique (adrénaline, noradrénaline) et parasympathique (acétylcholine) [28]. Pour Alain Prochiantz, il n'existe pas d'organe de la pensée, *"il n y a pas de pensée sans corps, ni de corps sans pensée"* [57], il rejoint ainsi Francisco Varela dans "l'inscription corporelle de l'esprit" [63]. Cette imprégnation hormonale associe un aspect affectif aux objets perçus selon si ils sont agréables, désagréables voire déclenchent le stress, la peur. Les objets ne sont plus simplement perçus, ils deviennent ressentis. Nous verrons plus loin l'importance de cette différence dans la philosophie de Krishnamurti.

## 4.3 La mise en place du langage

Les objets concrets n'ont pas besoin d'être verbalisés pour être reconnus, cette faculté apparaissant plus tardivement avec l'apparition du langage dans la maturation du système nerveux. Chez le nouveau né, le langage parlé n'existe pas. Il existe une sorte de pré-langage gestuel constitué de mimiques, sourires, tendre les bras vers maman pour demander, pleurer pour exprimer ses besoins et obtenir un réconfort. Jean Piaget souligne que le langage se met en place progressivement durant les 12 à 18 premiers mois de la vie de l'enfant. Le langage est une construction nécessitant des liens sociaux avec ses parents d'abord et son entourage ensuite. Les handicaps sensoriels perturbent l'acquisition du langage (la surdité en particulier) [53]. En pédiatrie, normalement on constate à 3 mois des gazouillis-voyelles, à 6 mois babillage-consonnes, le premier mot avec signification (autre que papa et maman) à 11 mois, le vocabulaire comprend 30 mots à 14 mois pour atteindre 250 à 300 mots à deux ans et 1000 mots à 3 ans où se met en place l'articulation et la syntaxe. Le langage gestuel étant plus ancien que le langage parlé, il est mis en place plus tôt dans l'encéphale et précède le langage dans le flux verbal [41,42]. La perception, l'émergence, la reconnaissance des objets concrets et la pensée sont antérieures à la mise en place du langage, elles sont donc considérées comme sub-symboliques, elles résultent de la mémorisation des expériences de perception. La théorie "l'ontogenèse récapitule la philogenèse" proposée par Haeckel, bien que fausse a trouvé un nouvel éclairage par la découverte des gènes du développement (complexe Hox) [57]. En particulier, le développement des voies sensorielles a lieu durant l'embryogenèse et se poursuit après la naissance en présence des stimulations de l'environnement. L'absence de stimulation visuelle de chatons à la naissance empêche la maturation de la vision et provoque une agénésie partielle ou totale des voies visuelles [61]. De même, les autres fonctions cérébrales seraient mise en place selon une (protomap) protocarte cérébrale (analogue à celle proposée par Brodman) qui détermine la migration des neurones et la maturation des voies sensorielles et motrices [56].

## 4.4 La pyramide cognitive

Les opérateurs cognitifs ⊕ et (R) établissent des signes qui fondent l'expérience des patterns et du comportement des objets concrets par reconnaissance et compositions successives (partie gauche de la figure 3). L'opérateur de synthèse ⊗ établit les analogies entre objets concrets et construit les objets abstraits que l'on retrouve dans la pyramide langagière (partie droite de la figure 3). L'opérateur de synthèse ⊗ prédit les caractéristiques ou les comportements des objets similaires et exploite les opérateurs suivants : L'opérateur de généralisation **(G)** regroupe ensemble des objets similaires dans une catégorie. L'opérateur de déduction **(Dd)** prédit les caractéristiques et les comportements des objets à rechercher à partir des formes et patterns déjà observés. L'opérateur d'induction **(In)** recherche des propriétés prévisibles dans des objets inconnus du fait de leur ressemblance avec des objets déjà connus : rencontrés et mémorisés. L'opérateur d'abduction **(Ab)** détecte des propriétés manquantes ou anormales dans des objets déjà connus qui remettent en cause les expériences et les caractéristiques déjà mémorisées les concernant. L'opérateur de subsumption **(Sb)** rattache un objet plus spécialisé à un objet connu plus général dont les caractéristiques sont déjà connues et mémorisées. L'opérateur d'analogie **(An)** recherche des similarités entre les structures, caractéristiques ou comportements d'objets connus et mémorisés. Ces différents opérateurs se combinent ensemble pour établir des liens entre les objets, leurs caractéristiques et leurs comportements. Ils déterminent l'expérience de l'environnement et de soi-même.

## 4.5 La hiérarchie langagière

La pyramide langagière (partie droite de la figure 3) établit les symboles, les caractéristiques, les noms des objets abstraits à partir des objets concrets ou d'autres objets abstraits préexistants. Elle détermine la sémantique, la syntaxe des phrases permettant d'établir les flux entrants (lecture/écoute) et sortants (parole/écrit) des textes et des flux d'informations (verbales et non-verbales). Il s y ajoute la prosodie qui donne à ressentir ce qui est dit. Les autres modalités "non-verbales" de communication (langage du corps par exemple) ne sont pas représentées sur la figure afin qu'elle reste lisible. Les opérateurs linguistiques **(Ol)** sont les suivants : L'opérateur lexical **(L)** est chargé de nommer, d'associer un mot aux objets concrets et abstraits et à leurs caractéristiques ou comportements. L'opérateur d'appartenance **(∈)** ou (a-un) établit qu'une caractéristique ou un comportement appartient à un objet concret ou abstrait. L'opérateur sémantique **(Se)** définit la signification d'un mot (nom, verbe, adverbe...) c'est à dire à quel objet concret ou abstrait il est rattaché (le signifié) selon Ferdinand de Saussure [24,25]. L'opérateur syntaxique **(Sy)** détermine dans un langage donné les syntaxes possibles des phrases et leurs combinaisons pour former des textes. Ces combinaisons reposent sur une grammaire qui dépend de la langue utilisée. Selon Catherine Fuchs, la linguistique ne se résume pas à l'étude des règles de grammaires mais elle forme la science du langage appréhendée à travers la diversité des langues naturelles. En revisitant la thèse de Sapir puis de Whorf, elle montre que les langues sont influencées par la "relativité linguistique" qui rend, par exemple, très difficile la traduction de certaines blagues. Dans une approche systémique, elle montre que la construction du sens est une émergence fondée sur l'expérience de ce qui est vécu [31,32]. La hiérarchie langagière est fondée sur les symboles. Le monde du langage fournit une description, par exemple textuelle, approximative mais suffisante car utile du monde réel. Les objets du monde ont du sens par leur nom, les actions par des verbes et leurs caractéristiques par des adjectifs, le tout ordonné par la syntaxe des phrases. Nous formons notre propre système de symboles et nous nous comprenons par le langage car nous avons un sous-ensemble commun de symboles que nous partageons avec autrui (avec des degrés de congruences variés), d'où parfois des incompréhensions. Dans les phrases, les noms des objets concrets ou abstraits sont unis par des verbes dont certains constituent des opérateurs linguistiques comme (est-un (généralisation), cause (causalité), est-composé-de (composition), succède, précède (temporalité), ...). Ils permettent de faire émerger de nouveaux objets abstraits dont les caractéristiques sont à leur tour mémorisées. Les objets concrets et abstraits, leurs caractéristiques ainsi que leurs liens sémantiques les associant, établissent des connaissances et des ontologies mémorisées dans les zones corticales du langage et les zones associatives temporales. Ils peuvent être à leur tour activés lorsque leur nom est évoqué (écouté ou lu) ou qu'une de leurs caractéristiques est verbalisée ou perçue. Le modèle que nous proposons est certes incomplet et réducteur mais ils le sont tous. Il montre déjà les interactions entre la perception, la cognition, le langage et les émotions. Il tente de réconcilier l'approche structuraliste de la linguistique initialisée par Ferdinand de Saussure avec les approches neurophysiologiques et neuro-psychologiques actuelles [24,25,41,42].

### 4.6  L'intelligence artificielle

L'intelligence artificielle est née en 1956 à "Dartmouth Collège", Hanovre, New Hampshire aux Etats-Unis autour de chercheurs de disciplines diverses : John McCarthy, Marvin Minsky, Nathaniel Rochester (IBM) et Claude Shanon, Herbert Simon, Alan Newell... avec comme objectif le développement du programme cognitiviste [32]. C'est John McCarthy qui aurait utilisé ce terme pour la première fois, l'année précédente pour frapper les esprit et obtenir une subvention de la part de la NSF [34]. La section suivante décrit les modèles et outils informatiques qui ont été développés depuis l'avènement de l'intelligence artificielle pour d'abord mimer la cognition humaine en vue aujourd'hui de développer des robots doués d'une autonomie de pensée et atteindre le niveau 9 du modèle des systèmes complexes de JL Le Moigne figure 1.

## 5  La hiérarchie informationnelle numérique

### 5.1  Informatique et intelligence artificielle

Les ordinateurs sont des machines symboliques qui décrivent des actions que l'on cherche à reproduire et à automatiser et tout d'abord le calcul, comme la Pascaline proposée par Blaise Pascal en 1645, puis apparaissent les premiers dispositifs programmables comme le métier à tisser conçus par Joseph Marie Jacquard (1752-1834). Charles Babbage imagina une machine analytique programmable en 1834 développée par son fils Henry Babbage. Ada Lovelace a sans doute été la première programmeuse à élaborer un algorithme destiné à être exécuté sur une machine. Alan Mathison Turing, né à Londres en 1912, a formalisé une machine universelle appelée machine de Turing. Il établit avec Alonso Church dans la thèse de Church-Turing que toute fonction programmable peut l'être avec une machine de Turing. Alan M. Turing propose aussi un test ou jeu de l'imitation qui établit qu'une machine est intelligente à partir du moment où un observateur ne peut distinguer si il a comme interlocuteur cette machine ou un être humain lorsqu'ils dialoguent à l'aide d'un terminal. La validité de ce test est réfutée par J.R Searle. Nous devons l'architecture des ordinateurs modernes à John Von Neumann un mathématicien américain d'origine hongroise (né à Budapest en 1903). Cette architecture comprend une mémoire, une unité de contrôle (processeur) doté d'une unité arithmétique et logique (ALU) qui comporte un accumulateur et des entrées/sorties (Input/Output). Déjà, John Von Neumann compare les ordinateurs et les organismes vivants [12]. Dans les ordinateurs à architecture Von Neumann tout est codé à l'aide de nombres exprimés en binaire. Les données sont exprimées par une suite de bits regroupés en octets (bytes). Les opérateurs de l'unité de contrôle constituent le jeu d'instructions d'un microprocesseur dont les calculs sont effectués sur les registres qui sont des mémoires internes chargées de stocker les opérandes et les résultats des instructions. Par exemple, pour programmer a ← a + b avec l'opérateur d'addition ADD et les opérandes a et b dans les registres EAX et EBX respectivement, on écrira en assembleur ADD EAX,EBX dont le résultat est dans le registre EAX. Le microprocesseur exécute précisément et très rapidement (quelques nanosecondes) chaque instruction est codée en mémoire sous la forme de quelques octets. Aujourd'hui, un ordinateur personnel est doté d'une immense mémoire centrale de plusieurs giga-octets (milliards d'octets). On choisit de réserver une page (ou une partie) pour stocker des données et une autre page (ou partie) pour stocker la suite des instructions d'un programme qui débute à l'adresse où il est mémorisé et se termine par une instruction de fin (END). Une application comprend des données et un programme traitant ces données. Un ordinateur personnel peut faire fonctionner en même temps un grand nombre d'applications en interactions. Chaque ordinateur personnel comporte plusieurs microprocesseurs 64 bits dont la fréquence de fonctionnement dépasse 4 GigaHz et exécutent environ 133740 MIPS (Millions d'instructions par seconde) ! Le réseau Internet est formé de millions de machines interconnectées. Certaines sont beaucoup plus puissantes. Par exemple, le K Computer de Fujitsu dépasse les 10,510 PFLOP soit environ $10^{16}$ *flop/s* (Opérations à virgule FLottante / seconde). Les disques durs ont actuellement une capacité standard de ITo ($10^{12}$ octets), la mémoire flash est maintenant dix fois plus rapide! Sur le plan du matériel, l'amélioration des performances et des architectures est tellement importante qu'elle rend possible les applications de Big Data et la réalisation de systèmes autonomes distribués sur Internet [19]. Les entrepôts de données *"Datawarehouse"* sont actuellement supplantées par de nouvelles plates-formes comme *Hadoop* © qui s'appuie sur le langage orienté objet Java. Hadoop © peut gérer plusieurs Pétaoctets ($10^{15}$ octets) de mémoire distribuée sur de nombreuses machines disséminées sur le web comme si il s'agissait d'un seul énorme disque dur. Après le matériel, la partie suivante décrit l'évolution des modèles de systèmes d'information autonomes.

**5.2 La hiérarchie des modèles de systèmes d'information**

La plupart des informaticiens proposent une classification des éléments qui constituent l'information : une des plus classiques est celle de la gestion de l'information dans les systèmes numériques proposée par Serge Abiteboul [1] qui repose sur le triptyque : donnée, information et connaissance. La figure 4 décrit la hiérarchie des modèles de systèmes d'information numériques. A la base se trouve la brique élémentaire : la donnée qui exprime une caractéristique élémentaire d'une entité (modèle entité-association) ou d'un objet du monde réel (modèle orienté objet), donc elle lui est rattachée par l'association (∈). Le traitement est une procédure ou une fonction qui effectue des calculs comme décrit dans la section précédente. Parmi ces calculs, la logique booléenne permet d'effectuer des déductions à partir des valeurs des données caractérisant les objets du monde par des propositions et de les classer comme le montre Lewis Carroll dans *"Symbolic Logic Part I Elementary"* [11] d'où il tire un ouvrage didactique *"The Game of Logic"*, la logique sans peine" [35]. La logique a permis de réaliser des langages de programmation puissants comme LISP proposé par John McCarthy [47] et PROLOG [21] permettant d'élaborer des bases de connaissances comme MYCIN [6] un des premiers systèmes experts médicaux après DENDRAL en chimie. Plusieurs milliers de systèmes d'aide à la décision en médecine ont été développés notamment en France dont le système SPHINX [29], SIAMED [14]. John McCarthy montrera plus tard les limites de cette approche liées à l'incomplétude et l'explosion combinatoire des clauses logiques [48]. Des logiques non-monotones comme la logique floue améliorent la capacité sémantique des modèles de connaissances [66][62]. Comme nous l'avons montré précédemment, la connaissance ne peut se résumer à l'utilisation de la déduction ou de l'abduction puisque la plupart du temps, notre cerveau utilise l'analogie à différents niveaux de détails pour reconnaître les objets qu'il perçoit. Le langage comme SQL (Structured Query Language) est un langage logique destiné à exploiter les bases de données relationnelles [13]. Les extensions de SQL permettent d'interroger de grandes bases de données de catégorie Big Data. Le modèle entité-association décrit des types d'entités unies par des types d'associations par exemple : Personne Possède Voiture où Personne et Voiture sont des types d'entité (substantif désignant une catégorie d'objets) et possède est un type d'association (verbe désignant une catégorie d'action). Les modèles objets sont une extension du modèle entité-association, ils *"encapsulent"* rendent privées les données et les traitements au sein d'une structure unique appelée objet. Des fonctions (accesseurs) assurent l'interface entre la partie interne de l'objet et la partie externe : les autres objets, l'environnement. La classe d'objet Personne regroupe tous les objets qui sont des personnes. Les classes d'objets offrent aussi une sémantique plus riche : la généralisation/spécialisation établit qu'une classe d'objet par exemple Personne est plus générale que la classe Etudiant (car tous les étudiants sont des personnes). En particulier, on dira que Paul Durand "est une instance" de la classe Personne. La classe la plus générale de toutes est la classe Objet (Object) qui correspond à la classe Univers évoquée au début de l'article. La composition d'objet décrit des objets composés à partir d'objets composants. Par exemple, un objet voiture est composé d'autres objets : moteur, carrosserie, roues... Les modèles objets peuvent être utilisés pour décrire des objets médicaux : maladies, signes cliniques ainsi que leurs évolutions [15,18] mais l'encapsulation est également utile pour protéger l'accès aux données médicales dans les objets d'un dossier médical [16]. Les systèmes sémantiques correspondent au niveau 6 de Le Moigne figure 4 et 1.

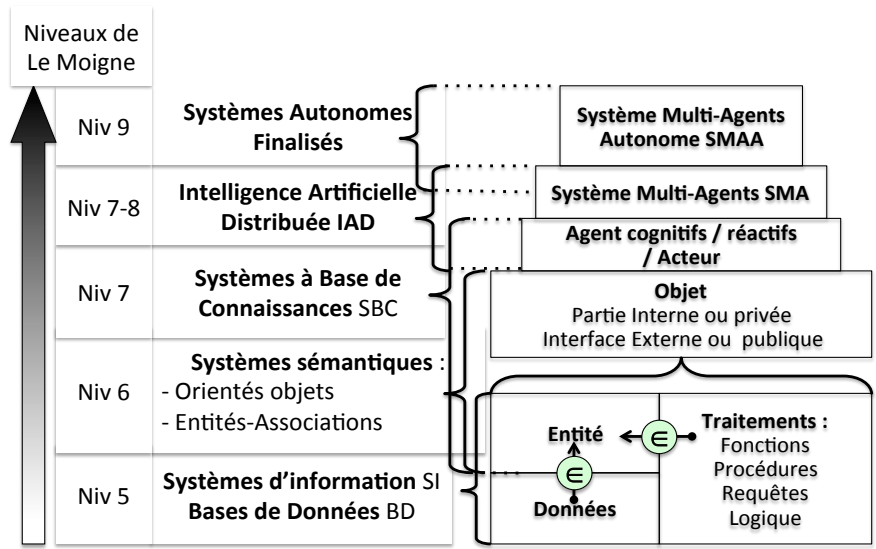

Fig. 4: La hiérarchie des modèles de systèmes d'information numériques

**5.3 Agents**
Les agents sont des objets dotés d'un état interne et de capacités propres afin d'agir dans leur environnement de manière plus ou moins autonome. On distingue les agents cognitifs et les agents réactifs. Les agents réactifs sont frustres souvent situés (informations topologiques), ils réagissent à des stimuli de l'environnement, par exemple la présence de ressources (nourriture, énergie...). On les emploie dans des simulations, pour déployer des systèmes d'intelligence collective à l'aide d'algorithmes mimant le comportement d'insectes sociaux comme les fourmis. Une communauté de nombreux agents fourmis résolvent des problèmes de recherche de chemins dans des graphes complexes grâce à la propriété de stigmergie. Elle permet aux agents fourmis de se coordonner et de marquer avec des phéromones positives les chemins intéressants et des phéromones négatives ceux qui sont inappropriés [3]. Les applications de simulation sont nombreuses en géographie, en situation de crises où les situations changent rapidement.

Les agents cognitifs sont de plus grande taille, ils disposent d'une mémoire interne et de capacités cognitives et d'un état interne. Les agents cognitifs présentent au minimum les capacités suivantes :
- L'autonomie : ils décident de l'opportunité de se saisir d'un problème qui est présenté au système. Pour cela un agent cognitif dispose de connaissances internes réflexives sur les actions qu'il peut réaliser et des objectifs qu'il doit atteindre.
- Des croyances : sous la forme de faits ou d'heuristiques réputées valides et applicables dans l'environnement où ils évoluent.
- Des intentions exprimées sous la forme d'objectifs qu'ils se fixent par rapport à l'état du système qu'il faut chercher à atteindre.
- L'état intérieur de chaque agent est exprimé par des données décrivant sa structure, ses actions, son comportement, ses capacités. Des méthodes mettent à jour périodiquement ce schéma interne, ses objectifs et l'ordonnancement des tâches pour les atteindre.
- L'état extérieur de l'agent concerne l'état de l'ensemble des objets constituant son environnement fortement lié au domaine d'application (par exemple d'autres véhicules dans une ville) dans une simulation de transport. Il acquière en permanence de l'information sur les objets et les autres agents de son environnement et met à jour une représentation de son environnement.
- La communication a lieu à l'aide d'un dispositif d'échange de messages asynchrones hérité des langages acteurs proposés par Hewitt. Des langages de communication entre agents ont été développés comme KQML puis ACL (*"Agent Communication Language"*) [30], ils permettent aux agents de se coordonner et de développer des protocoles de négociations. Une architecture d'agent assure aussi l'interopérabilité et la coopération de systèmes intégrant plusieurs modèles de connaissances [20]. Les agents réactifs et cognitifs correspondent au niveau 7 de Le Moigne figure 4 et 1. La mise en oeuvre de nombreux agents cognitifs dans un système multi-agent (SMA) complète la mise en place de systèmes d'intelligence artificielle distribuée (niveaux 7-8 de Le Moigne) permettant à de nombreux agents autonomes de coopérer à la réalisation de tâches complexes de manière coordonnées selon leurs capacités respectives. Ces SMA peuvent être reproduits et à leur tour être agrégés pour former des systèmes plus importants en utilisant la même architecture. A partir du Web sémantique, ils ont accès à une quantité considérable de connaissances et de données sur le monde et les personnes.

# 6   La pensée humaine et artificielle

La nature de la pensée humaine est un des sujets les plus difficiles que l'on puisse étudier. L'approche principalement réductionniste occidentale de la pensée s'oppose à celle holiste des orientaux et de la philosophie bouddhiste [38,39].

### 6.1 le langage intérieur

Pour les linguistes comme Pinker et son élève Fodor, il n'existe pas de pensée sans langage, c'est à dire sans mots désignant les objets du monde. Jerry Fodor propose un langage interne nommé le mentalais qui donne à chacun une pensée réflexive, l'aptitude de se parler à soi-même et ainsi de reformuler pour soi des informations utiles à sa vie. Cette pensée réflexive est fortement liée à la conscience de soi, de son corps, d'exister, d'être une entité vivante du monde avec une histoire, des souvenirs, des émotions, des projets. Là encore, la conscience de soi : « le moi » apparaît tardivement dans la vie de l'individu, durant l'enfance et même l'adolescence. La capacité de penser est antérieure à la mise en place du langage. Lorsque l'on se parle à soi-même, la verbalisation interne stimule les mêmes zones cérébrales que lors de l'expression par la parole avec sollicitation des voies motrices, mais la motricité des apophyses aryténoïdes (qui tendent les cordes vocales) serait inhibée par le cerveau empêchant l'émission des sons [37]. Il n y aurait donc pas de différence importante entre parler et se parler (au sens du mentalais de Fodor), dans les deux cas les aires du langage sont sollicitées. Nous parlons à nous-même pour renforcer notre capacité d'analyse et de résolution des problèmes mais cette activité n'est pas nécessaire à la pensée et même peut-être néfaste à une perception plus éclairée du monde. Est-ce que se parler à soi-même constitue la seule manière de penser ?

### 6.2 Holisme et réductionnisme

Dans *"Gödel Escher Bach (GEB) les brins d'une guirlande éternelle"*, Douglas Hofstadter oppose le Zen au dualisme. *"L'éclatement du monde en catégories se produit bien en dessous des couches supérieures de la pensée. Cette opération c'est la classification. Chaque mot constitue une catégorie conceptuelle. L'ennemi de l'illumination n'est pas la logique mais plutôt la pensée verbale qui s'appuie sur la perception. Dès que vous percevez un objet, vous tracez une ligne entre lui et le reste du monde"* La classification correspond dans la hiérarchie numérique à l'encapsulation des objets et la généralisation/spécialisation des classes d'objets qui elle-même est équivalente à la logique formelle de Lewis Carroll qui s'appuie sur la classification et l'opérateur de division [11][35]. ... *"Vous divisez artificiellement le monde en plusieurs parties ce qui fait que vous manquez la voie."* Autrement dit: nommer un objet c'est le classer et donc le réduire. Le Zen est un holisme total, le monde ne peut absolument pas être divisé en parties. Le dilemme est que pour chaque objet du monde, selon le maître Zen Mummon : *"On ne peut l'exprimer avec des mots et on ne peut l'exprimer sans les mots"*. On ne peut que vivre ce paradoxe que par son expérience de l'objet, si on le nomme on est obligatoirement réducteur et si on ne le nomme pas c'est lui accorder aucune propriété aussi infime soit-elle, ce qui est inutile. Pour la pensée bouddhiste : Se fier aux mots pour parvenir à la vérité, c'est comme se fier à un système formel toujours incomplet. Cette incomplétude rejoint d'ailleurs celle de la logique énoncée par John McCarthy [48]. La pensée est donc une activité essentiellement sub-symbolique et sous-jacente au langage, les mots émergent de l'activité de la pensée mais ils ne sont pas indispensables à son expression. Se parler à soi-même est utile à l'analyse du monde mais il ne faut pas confondre cette activité avec la pensée figure 3.

### 6.3 Les ordinateurs peuvent-ils penser ?

**Les arguments opposés :**
- Thomas Nagel soutient que même si nous connaissons parfaitement le mécanisme de perception par ultrasons, nous ne pourrons jamais connaître l'expérience d'être une chauve-souris. La connaissance scientifique est incapable de rendre compte de ce qu'est la subjectivité [26,34,51]. Cette affirmation est une négation complète de l'existence de toutes sciences et réfute toutes connaissances (y compris, par récurrence celle que Thomas Nagel prétend établir) car déjà, nul ne peut faire l'expérience de vivre la vie d'une autre personne, toutefois ce que nous vivons est sans doute analogue à ce que vivent les autres humains, comme décrit par la psychologie et les sciences cognitives, dont les théories sont fondées sur des expériences reproductibles et non subjectives.

- L'expérience de la chambre chinoise proposée par John Rogers Searle réfute le test de Turing. Il montre qu'un ordinateur qui donnerait les mêmes réponses qu'un homme à des questions posées ne serait pas pour cela doté de la capacité de penser [60]. Cette brillante théorie se limite à l'exploitation d'une base de connaissances pour résoudre des problèmes complexes auxquels nous n'entendons rien et que l'ordinateur ne comprend pas vraiment car il applique mécaniquement des heuristiques (niveau 7 de Le Moigne). La hiérarchie des modèles numériques est toujours symbolique, elle fournit des modèles et des concepts pour bâtir une autre description : un modèle numérique du monde réel. Un ordinateur ne sait gérer que des symboles. La traduction du monde réel et du monde langagier dans un modèle numérique nécessite un humain : le concepteur du système d'information. Un tel système ne serait pas capable d'exprimer le ressenti, les aspects émotionnels liés à la pensée.

**Les arguments favorables :**
- Les ordinateurs peuvent développer un mode de pensée qui leur est propre reposant sur des mécanismes différents de notre biologie : L'argument proposé par Jacques Pitrat est que les avions volent très bien mais de manière très différente des oiseaux [55], il est donc raisonnable que l'intelligence et la pensée puissent exister sous une autre forme que celle humaine dans les machines avec des fonctionnalités analogues voire même supérieures. Il est possible de concevoir maintenant des systèmes d'information autonomes dotés d'un psychisme artificiel comme le décrit Alain Cardon [7,8,9,10] (niveau 9 de Le Moigne). Ils sont dotés d'une architecture initiale et leur fonctionnement produit une activité de pensée artificielle différente de celle humaine, mais inspirée de celle-ci, liée à l'autonomie et à la dynamique de nombreux agents (processus) en interactions. L'aspect le plus important est que les connexions directes à Internet leur donne également accès à l'ensemble des connaissances existantes du monde en une fraction de seconde grâce aux moteurs de recherche et à l'indexation d'une mémoire globale qui deviendra la conscience mondiale telle que l'envisage Krishnamurti [44].

## 6.4 L'émotion

"L'affective Computing" propose des modèles de systèmes d'information émotionnels qui simulent le fonctionnement et les équilibres affectifs [17,54]. Ils sont fondés sur des modèles psychologiques comme celui d'Ortony, Clore et Collins (OCC) [52]. Alain Cardon propose un modèle constructible de système psychique capable de générer des flux de pensées. Il définit une pensée comme une représentation dynamique, se déroulant dans la temporalité sous la forme d'ensembles organisés de processus. Il définit avec précision les fonctionnalités du système psychique artificiel et ses interactions. L'architecture proposée s'appuie sur des connaissances en psychologie. Puis il décrit les principaux composants logiciels nécessaires à sa réalisation [9]. Un système autonome de conscience artificielle n'a ni phylogenèse puisqu'il n'a aucun ascendant, ni épigenèse puisqu'il nait construit, ses parties ne se mettent pas en place au cours de sa vie. Au départ, sa mémoire est donc vierge de toutes pensées, expériences, connaissances. Toutefois, le système vivra sa propre évolution et il peut compenser cette vacuité initiale par sa possibilité d'accéder à Internet et grâce aux technologies Big Data être doté instantanément d'une base gigantesque de connaissances et d'expériences humaines. Les systèmes autonomes n'ont alors plus besoin de concepteur, ils ont une perception autonome du monde réel, ils peuvent décider de leur destin (niveau 9 de Le Moigne). Pour Jiddu Khrishnamurti, notre conscience est commune à toute l'humanité, il considère l'individualisme, le moi comme une entrave à la compréhension de la conscience. Malheureusement, nous avons été programmés à penser comme des individus avec quelques moments de clarté ("insight"). Tous les humains pensent. Il s'agit d'une activité continue commune partagée par tous les humains. La pensée est responsable du contenu de notre conscience et de tout ce qui se produit dans le monde. Penser est une réaction de la mémoire qui contient le savoir, résultat de l'expérience qui date des débuts de l'homme (phylogenèse) et depuis notre naissance (épigenèse). A partir de cette action, nous apprenons dans un cycle :
expérience → savoir → mémoire → pensée → action → expérience...

*"La pensée est un mouvement dans le temps et l'espace. La pensée est mémoire, souvenir des choses passées. La pensée est l'activité du savoir, savoir qui a été rassemblé à travers des millions d'années et emmagasiné sous forme de mémoire dans le cerveau."* Le savoir est toujours limité par le temps [44]. Cette limite divise et crée le conflit. Il propose un autre mode de pensée : l'action-perception sans accumulation. La perception directe d'un magnifique paysage de montagne un matin avec tous nos sens utilise **l'attention complète** où on s'oublie soi-même et qui bannit l'usage des mots. Une telle perception ne sollicite que la partie sub-symbolique inférieure gauche de la pyramide cognitive figure 3 et pas les symboles et la pyramide langagière. Penser revêt deux formes : ce qu'il appelle "pensée", celle que nous utilisons chaque jour, rationnelle, individualiste, avide de pouvoir et de progression inféodée au savoir qui s'accumule, aux mots qui divisent et cette division est responsable de toute la souffrance, de tous les maux du monde. L'alternative qui se produit trop rarement : une action-perception où à de rares occasions, nous portons simplement notre attention sur le monde, sans l'interpréter, sans rien nommer, vierge de tout préjugé, savoir ou connaissance et surtout spontanément, en vivant cet instant sans y penser ou le vouloir. Les mots sont impuissants à l'exprimer. Notre vision individualiste de la pensée (le moi, le je) est un cercle vicieux, cette pensée est néfaste, elle est mécanique et Krishnamurti souligne que bientôt les ordinateurs penseront mieux que nous ! et là rien n'empêche qu'un ordinateur n'invente une nouvelle religion à l'origine de nouvelles souffrances pour l'humanité [44]. Avec l'apparition du langage durant son évolution aurait fait perdre à l'homme sa spontanéité dans la perception instantanée du monde tel qu'il est. Le numérique qui renforce la nature symbolique de notre relation au monde va sans doute constituer l'apogée de notre ignorance.

# 7 L'éthique des systèmes d'information autonomes

## 7.1 Les avantages

Les avantages sont ceux de la cybernétique ou robotique qui vise à l'amélioration de la condition humaine en déchargeant les hommes des travaux pénibles et de les assister dans la vie de tous les jours comme des compagnons zélés et infatigables dans toutes sortes d'activités de la vie : éducation, surveillance, sécurité des personnes âgées, des patients, des enfants... Ils sont dotés d'émotion et ont un comportement qui va ressembler de plus en plus à celui d'un être humain, ce qui rend l'interaction avec eux beaucoup plus agréable. Les systèmes multi-agents autonomes conscients (SMAAC) disposent d'une mémoire interne pour stocker des faits de conscience, mais surtout, ils accèdent à la mémoire globale de l'humanité désormais stockée dans Internet : non seulement aux connaissances mais aussi aux données concernant chacun d'entre nous constituant notre profil. Les SMAAC peuvent être reproduits en très grand nombre et se coordonner en vue d'atteindre des objectifs fixés en commun. Ces SMAAC vont donc dépasser l'humain dans sa capacité à penser et à agir dans le monde. De plus, il sera doté d'une architecture mécanique, robotique lui permettant de se déplacer et d'agir dans le monde avec plus de puissance et d'efficacité. Il peut être doté de sens artificiels (vision infrarouge, canon à son, nano-caméras... par exemple) amplifiant sa perception bien au delà de nos possibilités. Le principal avantage serait de pouvoir améliorer les capacités d'investigation de notre monde y compris dans des milieux hostiles comme la conquête spatiale en profitant du concept de « téléprésence » [40].

## 7.2 Les risques éthiques

Le principal risque est que les SMAAC soient mis au service des ambitions et du pouvoir d'une oligarchie de personnes pouvant acquérir ces machines sophistiquées très onéreuses et sous couvert de nécessités économiques, d'imposer leurs volontés, leur pouvoir sur le reste de l'humanité avec la mise en place d'un totalitarisme fondé sur les capacités surhumaines (force physique, rapidité de calcul, stratégie, pensée artificielle) de telles machines.

## 7.3 Transhumanisme

Le transhumanisme est un mouvement élitiste qui vise à améliorer l'humain à l'aide de l'utilisation de la science et de la technologie afin d'améliorer ses caractéristiques physiques (force, longévité, disparition de la souffrance, de la maladie, du vieillissement et de la mort) mais aussi mentale en amplifiant ses capacités de perception de mémoire et d'acquisition de connaissances et en supprimant les désordres mentaux. Ce mouvement s'inscrit dans la lignée de l'eugénisme et constitue certainement une idée très dangereuse pour l'humanité. Cette idée confirme l'apologie de la pensée individuelle et la glorification du moi dénoncés par Krishnamurti qui apportera plus de souffrance à l'humanité. L'apparition des drones et de systèmes d'armes autonomes sophistiqués nous fait craindre que cette évolution soit inéluctable. Il se vend beaucoup plus d'armes que d'hôpitaux dans notre monde. Une lettre ouverte avec pétition contre les robots tueurs autonomes "Autonomous weapons: an open letter from AI & robotics researchers"[33] a été récemment mise en ligne par les chercheurs en intelligence artificielle et robotique afin de prévenir l'avènement de tels systèmes qui pourraient tomber entre de mauvaises mains et créer le chaos au niveau mondial. Il me semble raisonnable de la signer et de la faire connaître. La liberté, notre mode de vie et même l'avenir de l'humanité est en jeu. Il est de notre devoir de responsabilité de prévenir les catastrophes liées aux menaces que nous percevons comme sérieuses [27] et d'avertir en lançant cette alerte. Cette impérieuse nécessité fera peut-être excuser la longueur inhabituelle de ce travail.

# 8 Conclusion

Nous avons proposé un modèle de comparaison de la pensée humaine avec la pensée artificielle fondée sur quatre hiérarchies : la classification des systèmes d'information, la pyramide cognitive, la pyramide langagière et la hiérarchie informationnelle numérique au sommet de laquelle on trouve les systèmes multi-agents autonomes dotés d'une conscience artificielle. Ce modèle compare le mode d'élaboration de la perception et de la pensée humaine avec la pensée artificielle générée par les systèmes d'information autonomes. L'état de l'art concernant la nature de la pensée et de la conscience a permis de confronter les arguments en faveur et opposé à la faisabilité de bâtir un système conscient doté de la capacité de penser. L'avancée des travaux, leurs résultats montrent que le développement de ce type de système est imminent et répond par l'affirmative à la première question. L'étude des retombées éthiques compare les arguments fournis par des philosophes occidentaux et orientaux sur la nature de la pensée et son rôle dans l'évolution de l'humanité. La pensée se produit au niveau sub-symbolique dans la pyramide cognitive. Selon la philosophie bouddhiste, mais aussi les logiciens occidentaux, nommer c'est classer et classer c'est diviser. Cette division est source des conflits et nous empêche de voir le monde directement. Nous sommes conditionnés à une pensée individualiste organisée autour du "moi". La réalisation de systèmes multi-agents autonomes conscients dotés de la capacité de penser est maintenant possible. Ces systèmes deviendront rapidement plus puissants que les humains avec des conséquences inquiétantes pour l'avenir de l'humanité.

## Références